\documentclass[paper,notoc]{JHEP3}
\usepackage{epsfig,cite}
\usepackage{amsbsy} 
\usepackage{epsfig}
\usepackage{graphicx}

\newcommand\new{\newcommand}         % shorthand for \newcommand
\newcommand{\pthmax}{\ensuremath{p_{\mathrm{T}}^{\mathrm{H,max}}}}
\newcommand{\pthiggs}{\ensuremath{p_{\mathrm{T}}^{\mathrm{H}}}}
\newcommand{\ptjet}{\ensuremath{p_{\mathrm{T}}^{\mathrm{jet}}}}
\newcommand{\mh}{\ensuremath{m_\mathrm{H}}}
\newcommand{\mll}{\ensuremath{m_{\ell\ell}}}
\newcommand{\phill}{\ensuremath{\phi_{\ell\ell}}}
\newcommand{\ptmax}{\ensuremath{p_\mathrm{T,max}^{\ell}}}
\newcommand{\ptmaxcut}{\ensuremath{p_\mathrm{T,max}^{\ell,\mathrm{cut}}}}
\newcommand{\etmiss}{\ensuremath{E_{\mathrm{T}}^{\mathrm{miss}}}}

\newcommand{\accsigma}{\ensuremath{\sigma_{\mathrm{acc}}}}

\new{\emem}{{\ifmmode\mathrm{e}^-\else e$^-$\fi}}
\new{\epem}{{\ifmmode\mathrm{e}^+\else e$^+$\fi}}
\new{\zbo}  {{\ifmmode\mathrm{Z}\else Z\fi}}
\new{\wpm} {{\ifmmode\mathrm{W}^\pm\else W$^\pm$\fi}}
\new{\wbo} {{\ifmmode\mathrm{W}\else W\fi}}
\new{\epm} {{\ifmmode\mathrm{e^+e^-}\else $\mathrm{e^+e^-}$\fi}}
\new{\qq}  {{\ifmmode\mathrm{q}\else q\fi}}
\new{\qqb} {{\ifmmode\bar{\mathrm{q}}\else $\bar{\mathrm{q}}$\fi}}
\new{\tq}  {{\ifmmode\mathrm{t}\else t\fi}}
\new{\tqb} {{\ifmmode\bar{\mathrm{t}}\else $\bar{\mathrm{t}}$\fi}}
\new{\bq}  {{\ifmmode\mathrm{b}\else b\fi}}
\new{\bqb} {{\ifmmode\bar{\mathrm{b}}\else $\bar{\mathrm{b}}$\fi}}
\new{\ttbar}{\tq\tqb}
\new{\qqbar}{\qq\qqb}
\new{\gu}  {{\ifmmode\mathrm{g}\else g\fi}}
\new{\qqbarg}{\qq\qqb\gu}
\new{\pp}  {{\ifmmode\mathrm{p}\else p\fi}}

\new{\hh}  {{\ifmmode h\else $h$\fi}}
\new{\HH}  {{\ifmmode \mathrm{H}\else $\mathrm{H}$\fi}}
\new{\fe}  {{\ifmmode f\else $f$\fi}}
\new{\lp}  {{\ifmmode \ell\else $\ell$\fi}}
\new{\XX}  {{\ifmmode X\else $X$\fi}}
\new{\Vp}  {{\ifmmode V\else $V$\fi}}
\new{\Kzs} {{\ifmmode\mathrm{K}_\mathrm{S}^0\else $\mathrm{K}_\mathrm{S}^0$\fi}}
\new{\Kzl} {{\ifmmode\mathrm{K}_\mathrm{L}^0\else $\mathrm{K}_\mathrm{L}^0$\fi}}
\new{\Kp} {{\ifmmode\mathrm{K}\else $\mathrm{K}$\fi}}
\new{\ppHWW} {{\ifmmode\pp\pp\rightarrow\HH\rightarrow\wbo\wbo
                             \else $\pp\pp\rightarrow\HH\rightarrow\wbo\wbo$\fi}}
\new{\ppHWWlept} {{\ifmmode\pp\pp\rightarrow\HH\rightarrow\wbo\wbo\rightarrow\lp\nu\lp\nu
                             \else $\pp\pp\rightarrow\HH\rightarrow\wbo\wbo\rightarrow\lp\nu\lp\nu$\fi}}
\new{\ppHWWleptX} {{\ifmmode\pp\pp\rightarrow\HH+X\rightarrow\wbo\wbo+X\rightarrow\lp\nu\lp\nu+X
                             \else $\pp\pp\rightarrow\HH+X\rightarrow\wbo\wbo+X\rightarrow\lp\nu\lp\nu+X$\fi}}

\new{\HWWlept} {{\ifmmode\HH\rightarrow\wbo\wbo\rightarrow\lp\nu\lp\nu
                             \else $\HH\rightarrow\wbo\wbo\rightarrow\lp\nu\lp\nu$\fi}}
\new{\HWW} {{\ifmmode\HH\rightarrow\wbo\wbo
                             \else $\HH\rightarrow\wbo\wbo$\fi}}

\new{\WW} {{\ifmmode\wbo\wbo
                             \else $\wbo\wbo$\fi}}
                            
\new{\pptt}{{\ifmmode\pp\pp\rightarrow\ttbar
                             \else $\pp\pp\rightarrow\ttbar$\fi}}
\new{\ppWW}{{\ifmmode\pp\pp\rightarrow\wbo\wbo
                             \else $\pp\pp\rightarrow\wbo\wbo$\fi}}
\new{\Hgammagamma} {{\ifmmode\HH\rightarrow\gamma\gamma
                             \else $\HH\rightarrow\gamma\gamma$\fi}}

\new{\LEP}        {\mbox{\small\textsc{LEP}}}
\new{\LEPONE}     {\mbox{\small\textsc{LEP1}}}
\new{\LEPTWO}     {\mbox{\small\textsc{LEP2}}}
\new{\CERN}       {\mbox{\small\textsc{CERN}}}
\new{\ALEPH}      {\mbox{\small\textsc{ALEPH}}}
\new{\DELPHI}     {\mbox{\small\textsc{DELPHI}}}
\new{\LD}         {\mbox{\small\textsc{L3}}}
\new{\OPAL}       {\mbox{\small\textsc{OPAL}}}
\new{\SPS}        {\mbox{\small\textsc{SPS}}}
\new{\TEVATRON}   {\mbox{\small\textsc{TEVATRON}}}
\new{\LHC}        {\mbox{\small\textsc{LHC}}}
\new{\FERMILAB}   {\mbox{\small\textsc{FERMILAB}}}
\new{\CDF}        {\mbox{\small\textsc{CDF}}}
\new{\DZERO}      {\mbox{\small\textsc{D0}}}
\new{\CTEQ}        {\mbox{\small\textsc{CTEQ}}}
\new{\FNAL}        {\mbox{\small\textsc{FNAL}}}
\new{\ATLAS}        {\mbox{\small\textsc{ATLAS}}}
\new{\CMS}        {\mbox{\small\textsc{CMS}}}

\new{\eV}         {{\ifmmode {\mathrm{ eV}}\else ${\mathrm{ eV}}$\fi}}
\new{\MeV}        {{\ifmmode {\mathrm{ MeV}}\else ${\mathrm{ MeV}}$\fi}}
\new{\MeVc}       {{\ifmmode {\mathrm{ MeV}}/c\else ${\mathrm{ MeV}}/c$\fi}}
\new{\MeVcc}      {{\ifmmode {\mathrm{ MeV}}/c^2\else ${\mathrm{ MeV}}/c^2$\fi}}
\new{\GeV}        {{\ifmmode {\mathrm{ GeV}}\else ${\mathrm{ GeV}}$\fi}}
\new{\GeVc}       {{\ifmmode {\mathrm{ GeV}}/c\else ${\mathrm{GeV}}/c$\fi}}
\new{\GeVcc}      {{\ifmmode {\mathrm{ GeV}}/c^2\else ${\mathrm{GeV}}/c^2$\fi}}
\new{\TeV}        {{\ifmmode {\mathrm{ TeV}}\else ${\mathrm{ TeV}}$\fi}}

\new{\fb}        {{\ifmmode {\mathrm{ fb}}\else ${\mathrm{ fb}}$\fi}}
\new{\fbinv}   {{\ifmmode {\mathrm{ fb}^{-1}}\else ${\mathrm{ fb}^{-1}}$\fi}}
\new{\pb}        {{\ifmmode {\mathrm{ pb}}\else ${\mathrm{ pb}}$\fi}}
\new{\pbinv}   {{\ifmmode {\mathrm{ pb}^{-1}}\else ${\mathrm{ pb}^{-1}}$\fi}}

\new{\JS}         {\mbox{\small\textsc{JETSET}}}
\new{\HERWIG}         {\mbox{\small\textsc{HERWIG}}}
\new{\PYTHIA}         {\mbox{\small\textsc{PYTHIA}}}
\new{\AR}         {\mbox{\small\textsc{ARIADNE}}}
\new{\PY}         {\mbox{\small\textsc{PYTHIA}}}
\new{\JSv}        {\mbox{\small\textsc{JETSET\ 7.405}}}
\new{\HWo}        {\mbox{\small\textsc{HERWIG\ 5.8}}}
\new{\HWn}        {\mbox{\small\textsc{HERWIG\ 5.9}}}
\new{\ARv}        {\mbox{\small\textsc{ARIADNE\ 4.05}}}
\new{\PYv}        {\mbox{\small\textsc{PYTHIA\ 5.7}}}

\new{\fehip}        {\mbox{\small\textsc{FEHiP}}}
\new{\fewz}        {\mbox{\small\textsc{FEWZ}}}
\new{\hqt}        {\mbox{\small\textsc{HqT}}}
\new{\mcnlo}        {\mbox{\small\textsc{MC@NLO}}}
\new{\pvegas}        {\mbox{\small\textsc{PVEGAS}}}

\new{\Mz}         {{\ifmmode M_{\mathrm{ Z}}
                    \else $M_{\mathrm{ Z}}$\fi}}
\new{\Mzsq}       {{\ifmmode M^2_{\mathrm{ Z}}
                    \else $M^2_{\mathrm{ Z}}$\fi}}
\new{\Mw}         {{\ifmmode M_{\mathrm{ W}}
                    \else $M_{\mathrm{ W}}$\fi}}
                    
\new{\MH}         {{\ifmmode m_{\mathrm{ H}}
                    \else $m_{\mathrm{ H}}$\fi}}

\new{\as}[1]      {{\ifmmode\alpha^{#1}_s
                    \else$\alpha^{#1}_s$\fi}}
\new{\asx}[1]      {{\ifmmode a^{#1}_s
                    \else $a^{#1}_s$\fi}}
\new{\asb}[1]     {{\ifmmode\overline{\alpha}^{#1}_s
                    \else $\overline{\alpha}^{#1}_s$\fi}}
\new{\asmz}       {{\ifmmode\alpha_s(\Mzsq)
                    \else $\alpha_s(\Mzsq)$\fi}}
\new{\lqcd}       {{\ifmmode\Lambda_{\mathrm{ QCD}}
                    \else $\Lambda_{\mathrm{ QCD}}$\fi}}
\new{\lqcdsq}     {{\ifmmode\Lambda^2_{\mathrm{ QCD}}
                    \else $\Lambda^2_{\mathrm{ QCD}}$\fi}}
\new{\llla}       {{\ifmmode\Lambda_{\mathrm{ LLA}}
                    \else $\Lambda_{\mathrm{ LLA}}$\fi}} 
\new{\lmsbar}[1]  {{\ifmmode \Lambda^{(#1)}_{\overline{\mathrm{MS}}}
                    \else $\Lambda^{(#1)}_{\overline{\mathrm{MS}}}$\fi}}
\new{\lmsb}       {{\ifmmode \Lambda_{\overline{\mathrm{MS}}}
                    \else $\Lambda_{\overline{\mathrm{MS}}}$\fi}}
\new{\lmsbsq}     {{\ifmmode \Lambda^{2}_{\overline{\mathrm{MS}}}
                    \else $\Lambda^{2}_{\overline{\mathrm{MS}}}$\fi}}
\new{\pt}       {{\ifmmode p_{\mathrm{T}}
                    \else $p_{\mathrm{T}}$\fi}}
\new{\etmisscut}       {{\ifmmode E_{\mathrm{T}}^{\mathrm{miss,cut}}
                    \else $E_{\mathrm{T}}^{\mathrm{miss,cut}}$\fi}}
\new{\ptlmin}       {{\ifmmode p_{\mathrm{T}}^{\lp\mathrm{min}}
                    \else $p_{\mathrm{T}}^{\lp\mathrm{min}}$\fi}}
\new{\ptlmaxcut}       {{\ifmmode \mathrm{p}_{\mathrm{T,max}}
                    \else $\mathrm{p}_{\mathrm{t,max}}^{\mathrm{cut}}$\fi}} 
\new{\ptlep}       {{\ifmmode p_{\mathrm{T}}^{\mathrm{lepton}}
                    \else $p_{\mathrm{T}}^{\mathrm{lepton}}$\fi}}  
\new{\ptveto}       {{\ifmmode p_{\mathrm{T}}^{\mathrm{veto}}
                    \else $p_{\mathrm{T}}^{\mathrm{veto}}$\fi}}          
\new{\kt}       {{\ifmmode k_{\mathrm{T}}
                    \else $k_{\mathrm{T}}$\fi}}          

\title{\boldmath QCD radiation effects on the 
\HWWlept\ signal at the LHC}

\author{Charalampos Anastasiou\\
  Institute for Theoretical Physics, ETH Zurich,\\
  8093 Zurich, Switzerland\\
  E-mail: \email{babis@phys.ethz.ch}}
\author{G\"unther Dissertori\\
  Institute for Particle Physics, ETH Zurich,\\
  8093 Zurich, Switzerland\\
  E-mail: \email{dissertori@phys.ethz.ch}}
\author{Fabian St\"ockli\\
  Institute for Particle Physics, ETH Zurich,\\
  8093 Zurich, Switzerland\\
  E-mail: \email{fabstoec@phys.ethz.ch}}
\author{Bryan R. Webber\\
  Cavendish Laboratory\\
  Cambridge CB3 0HE, U.K.\\
  E-mail: \email{webber@hep.phy.cam.ac.uk}}

\abstract{ 
The discovery of a Standard Model Higgs boson is possible 
when experimental cuts are applied which increase the 
ratio of signal and background cross-sections. 
In this paper we study the \ppHWW~signal cross-section
at the LHC which requires a selection of Higgs bosons with 
small transverse momentum. 
We compare predictions for the efficiency of the 
experimental cuts  from 
a  NNLO QCD calculation, a calculation of the resummation of 
logarithms in the transverse momentum of the Higgs boson at NNLL, 
and the event generator \mcnlo. 
We also investigate the impact of various jet-algorithms, the underlying 
event and hadronization on the signal cross-section. 
} 
\preprint{{ETHZ-IPP/PR-2008-01}\\ {Cavendish-HEP-08/01}}

\begin{document}

\section{Introduction}
\label{sec:intro}
Physical  processes at collider experiments 
can be simulated using  flexible event generators such as 
\PYTHIA~\cite{pythia} and \HERWIG~\cite{herwig}.  
In this approach, the momenta of partons in hard scattering 
processes are distributed among hadrons using an approximate  
probabilistic algorithm for parton branching and hadronization.  
Traditionally, event generators compute the hard scattering partonic 
cross-section at leading order in fixed order perturbation theory 
which yields only a rough estimate.

Cross-sections for the hard interaction of three or  four 
particles can be computed routinely through next-to-leading order (NLO) in 
the $\alpha_s$ expansion. 
The results from NLO QCD calculations 
and parton shower event generators are often combined with an 
empiric method.  First the efficiency of the experimental cuts and normalized 
differential distributions are computed with parton shower event generators.  
Then, they are multiplied with the  result 
for the total cross-section from the NLO 
calculation~\footnote{Note that differential ``$K$-factors'' 
to reproduce bin integrated differential distributions at higher 
orders in perturbation theory  are also used~\cite{reweight1,reweight2}.}. 
However, it is now understood how to combine parton shower generators 
and NLO results with theoretically sound methods so that 
(i)  
leading logarithms  are resummed with the parton shower, and 
(ii)  all differential cross-sections are exactly accurate 
through NLO upon an expansion of the Sudakov factors in 
$\alpha_s$~\cite{mcnlo,matching2}.

Fixed order perturbative computations have now advanced beyond the next 
to leading  order, and there are two hadron collider processes which are 
known through next-to-next-to-leading order (NNLO) in  $\alpha_s$. 
The NNLO total cross-section of the Drell-Yan process~\cite{DYtotal1,Htotal1} 
is the most precise theoretical 
prediction for a hadron collider observable with a scale 
variation uncertainty of about $1\%$. 
The total cross-section for Higgs boson  production is 
also known at NNLO~\cite{Htotal1,Htotal2,Htotal3,Htotal4,Htotal5}. 
For the LHC, the NLO~\cite{Hnlo1,Hnlo2} 
and NNLO corrections are both important and
increase the LO result by about $70\%$ and $30\%$ respectively. 
The perturbative series converges slowly and the remaining 
theoretical uncertainty is about 
$\pm 10\%$~\cite{Hcouplings,Htotal1,Htotal2,Htotal3,Htotal4,Htotal5}. 

In Higgs boson~\cite{Hdiff1,fehip,Hdiff2} 
and electroweak gauge boson~\cite{DYrap1,DYrap2, Whip1, Whip2} 
production there exist novel differential cross-section calculations at 
NNLO~\cite{Hdiff1,fehip,DYrap1,DYrap2,Whip1,Whip2,Hdiff2}~\footnote{In non-hadronic collisions the state of the art at NNLO is fully differential 
cross-sections for $e^+e^-\to 3\,\mbox{jets}$~\cite{nnlo3jets1,nnlo3jets2,nnlo3jets3}.}. 
The cross-sections with arbitrary experimental cuts applied at the parton level 
can be  computed exactly  at this order in pertubration theory 
for the two processes. 
It is very instructive to compare the efficiencies of experimental 
cuts  from the newly available fully differential NNLO calculations, 
merged NLO and parton-shower calculations with \mcnlo, and simple 
leading order  event generators. 
This is valuable in order to estimate the inherent theoretical uncertainties 
of the above approaches. 

Such a  comparison can be made for the 
Drell-Yan process with the results from Refs.~\cite{Whip1,Whip2,Wacc}, as well as
discussed in Ref.~\cite{Whip1}.  
\mcnlo~and the NNLO Monte-Carlo \fewz~\cite{Whip1,Whip2} 
predict very similar 
experimental efficiencies for the entire kinematic range 
where the NNLO prediction retains its phenomenal scale variation  of 
about $1\%$. Significant differences, however, arise when the experimental 
cuts suppress contributions  from  the two-loop matrix elements.

A similar comparison~\cite{hppnnlogen} 
between \mcnlo~and the NNLO partonic Monte-Carlo 
\fehip~has been made for the Higgs boson diphoton signal at the LHC. The 
signal cross-section is known with a scale 
uncertainty of about $\pm 7\%$ at NNLO~\cite{fehip}. 
Besides a relatively 
large perturbative correction from NLO to NNLO of about $20\%$, the 
efficiency of the experimental cuts turns out to be very similar 
in \mcnlo~and NNLO~\cite{hppnnlogen}.

A challenging channel for the Higgs boson discovery at the LHC  is 
$\ppHWW$. This channel is contaminated by background processes, 
$\pptt$ and $\ppWW$, with much larger 
cross-sections. For a Higgs boson with a mass close to the 
\wbo-pair threshold, the cross-sections for all other discovery 
signals are suppressed. In this case, 
the $\ppHWWlept$ process becomes the only viable 
channel for the Higgs boson to be discovered at the LHC. 
It is thus indispensable to achieve a very good signal to background (S/B)
ratio.  An optimized selection of \wbo-pair events~\cite{hwwcuts} is then 
required. If appropriate 
cuts are applied, as in Refs.~\cite{newcuts1,newcuts2,newcuts3,newcuts4}, 
a discovery of a Standard Model Higgs boson 
with a mass close to the threshold should be achieved  with an 
integrated luminosity  of a few $\mathrm{fb}^{-1}$ at the LHC.

The \ppHWWlept~decay mode was recently implemented
in the NNLO Monte-Carlo \fehip~\cite{fehip} and a calculation of the 
cross-section with the experimental cuts of 
Ref.~\cite{newcuts1,newcuts2} was performed in Ref.~\cite{fehipWW}. 
The very good agreement of the \mcnlo~and NNLO calculations 
for the efficiency of the experimental cuts in the diphoton 
signal~\cite{hppnnlogen,fehip,mcnlo} does not guarantee
that  a similarly good agreement will be found in 
the $\HWW$ channel.  In this paper, we will compare the 
NNLO predictions from Ref.~\cite{fehipWW} 
with resummation calculations and event generators. 

In the \Hgammagamma~channel, the experimental 
cuts
select  events  ``democratically'', irrespectively of 
the transverse momentum of the jets associated with the Higgs boson 
production. The event selection in the \HWW~channel~\cite{newcuts2,newcuts3}  imposes  
an explicit jet-veto and other cuts which reject events with large 
transverse momentum \pthiggs~for the Higgs boson. 
This may turn out to be problematic for an agreement in experimental 
efficiencies between  \mcnlo~and NNLO for two reasons. 
First, the NNLO/NLO $K$-factor is sensitive to the 
cutoffs imposed on the \pthiggs~\cite{cataniveto,Hdiff1,fehip}; 
this effect is treated only in the parton shower approximation in \mcnlo.
Second, by selecting events with low \pthiggs, multiple gluon radiation 
effects which are not included in the fixed order NNLO 
calculation may be important.

The resummation through next-to-next-to-leading logarithms (NNLL) of
\pthiggs~is  now achieved~\cite{ptresum1,ptresum2,ptresum3}. 
The resummed spectrum, after matching to fixed order, 
integrates to the NNLO total cross-section. This calculation 
takes into account consistently  both multiple 
gluon radiation effects at low transverse momentum 
and fixed order high transverse momentum contributions.

A theoretical prediction for the signal cross-section 
\ppHWWlept~cannot be made directly 
from the \pthiggs~spectrum, since 
the experimental cuts restrict many phase-space variables.
The cross-section must therefore be computed using  fully differential 
Monte-Carlo programs.
However, we will use the resummation calculation of the 
\pthiggs~spectrum~\cite{hqt,ptresum2,ptresum3} 
to validate fixed order Monte-Carlo's 
and parton shower event generators in the low \pthiggs~kinematic  
region, which is favored by the experimental selection 
cuts.  

We first compare in Section~\ref{sec:comp1} the theoretical calculations  
from the NNLL resummation~\cite{hqt},  \mcnlo~\cite{mcnlo}
and NNLO~\cite{fehipWW} for the \pthiggs~spectrum.  
Then we compare in Section~\ref{sec:comp2} 
the \mcnlo~\cite{mcnlo}, \HERWIG~\cite{herwig}
and NNLO~\cite{fehipWW} Monte-Carlo predictions
for kinematic distributions of  
variables which are restricted by the experimental selection.\footnote{In
the case of \mcnlo\ and \HERWIG, we use a modified version of \HERWIG\ with
the correct decay angular correlations.}
We perform a similar comparison for the signal
cross-sections when all cuts are applied. 
We find a good agreement for the 
efficiencies of experimental cuts and normalized kinematic distributions. 
This gives us confidence that the 
selection of events for the signal 
cross-section does not invalidate the approximations in the used 
Monte-Carlo programs. 

In Section~\ref{sec:jetalgo} we study the sensitivity of the 
signal cross-section with all experimental cuts applied to the 
choice of jet algorithm. We find a mild change on the cross-section 
($\sim 6\%$) by using the SISCone or the  \kt-algorithm. 
We also study the effect of hadronization 
(using the model in \HERWIG~\cite{herwig})  
and of the underlying event (using the JIMMY model~\cite{jimmy}).
We find that proposed experimental cuts render the cross-section 
sensitive to these effects at the level of up to $5-10\%$. 
The changes to the cross-section are with opposite signs 
and  the combined effect is  rather mild.

\section{Integrating over the transverse momentum of the Higgs boson}
\label{sec:comp1}

\begin{figure}[h]
\begin{center}
\includegraphics[width=0.70\textwidth]{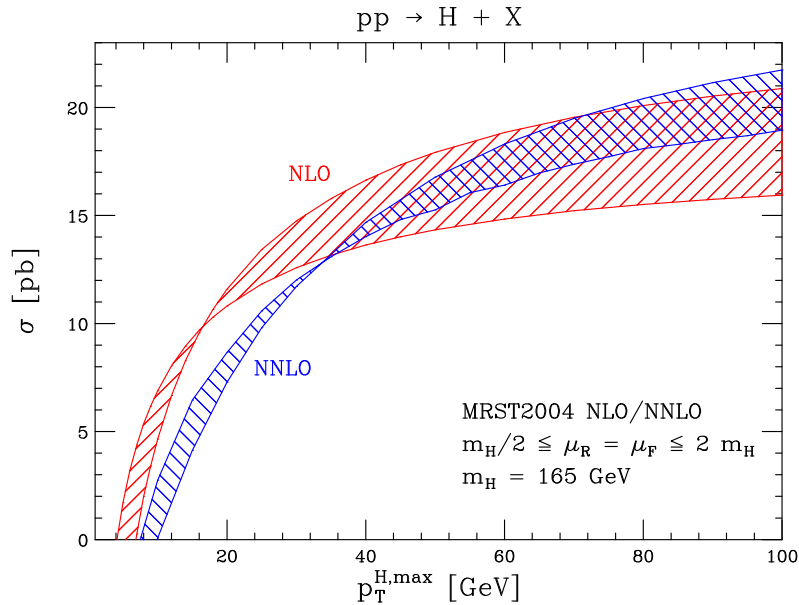}
\caption{Cumulative cross-section for the Higgs transverse momentum 
distribution at fixed NLO and NNLO \as~ expansion.}
\label{fig:higgspt_fixorder}
\end{center}
\end{figure}
In this Section we compare fixed order~\cite{fehip} and 
parton shower calculations~\cite{herwig,mcnlo} 
against the theoretical predictions for the 
transverse momentum spectrum of the Higgs boson from resummation~\cite{hqt}.  
Fixed order perturbation theory is invalid for  small values of \pthiggs.
Nevertheless, observables can be reliably 
computed if their definition allows for an integration over a 
sufficiently large range in \pthiggs~values.  
Here, we study the cumulative cross-section 
\begin{equation}
\sigma(\pthmax)=\int\limits_{0}^{\pthmax}\frac{\partial\sigma}{\partial\pthiggs}\;d\pthiggs\;.
\end{equation}
This observable mimics the effect of selection cuts 
with a cutoff on the maximum Higgs transverse momentum. 
Since in the transverse plane the Higgs boson balances 
the associated jet radiation, 
the cross-section with a jet-veto is similar to the cross-section with a 
veto on high values of \pthiggs.

\begin{figure}[h]
\begin{center}
\includegraphics[width=0.70\textwidth]{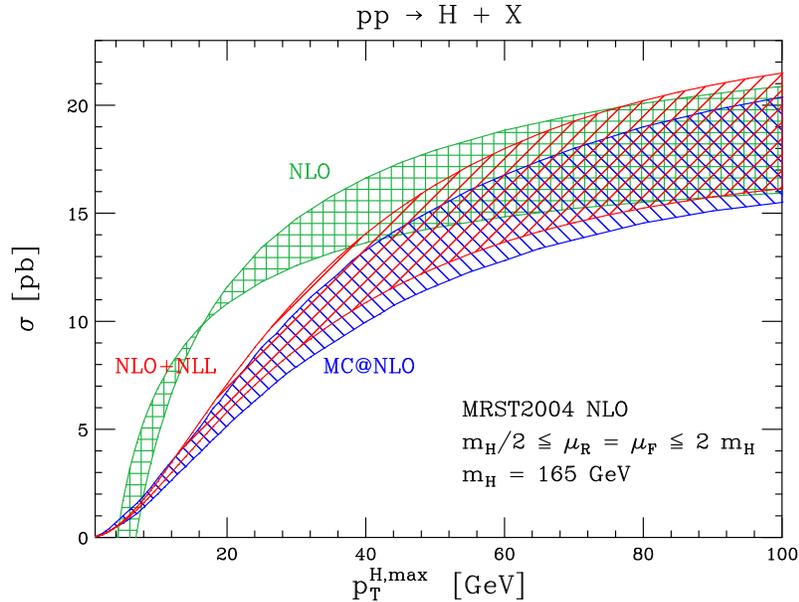}
\caption{Cumulative cross-section for the Higgs transverse momentum 
distribution with \mcnlo, NLL resummation~\cite{hqt} and at NLO.}
\label{fig:higgspt_nlo}
\end{center}
\end{figure}
In Fig.~\ref{fig:higgspt_fixorder} we plot 
the cross-section in the fixed NLO and NNLO \as~ expansion.  
As in all plots of this paper, we show the scale variation 
in the interval $\MH/2 < \mu_F =  \mu_R < 2 \MH$.  
The fixed order cross-sections at NLO and NNLO 
become negative and tend to infinity 
when the \pthiggs~cutoff is small (below $10 \, \GeV$ ). 
For such small values of \pthmax~perturbation theory breaks down. 
We observe that for larger cutoffs (above $40\, \GeV$ and up to 
$100 \, \GeV$) the
NLO and NNLO results are in very good agreement.  The NNLO cross-section 
increases  however faster than NLO with higher cutoffs leading to the 
known by $\sim 20\%$ larger NNLO result with respect to 
NLO for  the total cross-section~\cite{Htotal1,Htotal2,Htotal3}. 
The analyses in Ref.~\cite{newcuts1,newcuts2} show that a better 
S/B  ratio is achieved if stricter than 40 \GeV~cutoffs 
are used for the jet transverse momenta; for cutoff values 
in between $20\,\GeV$ and $40\,\GeV$ we observe large  perturbative 
corrections.  

In Fig.~\ref{fig:higgspt_nlo} we compare the fixed NLO result with the 
resummed NLL \pthiggs~spectrum from Ref.~\cite{hqt} and with 
\mcnlo~\cite{mcnlo} without hadronization and underlying event.  
We find that for $\pthiggs<  40\; \GeV$  the  parton shower or the NLL 
resummation change significantly the integrated NLO \pthiggs~distribution.
All results converge for higher 
and higher cutoffs \pthmax~and agree with each other for the fully 
inclusive cross-section. 
Notably, the \mcnlo~and the NLL resummation are in a rather 
good agreement with each other within the uncertainties 
from scale variation.

Differences between cross-section predictions 
with fixed order perturbation theory  and resummation  are 
expected to become smaller when the fixed order calculations are extended 
to higher orders. 
\begin{figure}[h]
\begin{center}
\includegraphics[width=0.70\textwidth]{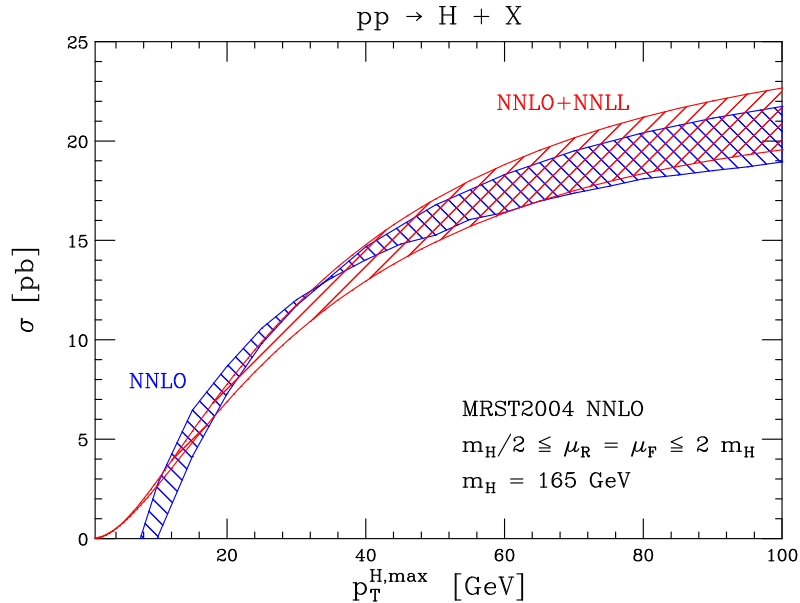}
\caption{Cumulative cross-section for the Higgs transverse momentum 
distribution at NNLO in fixed order and with NNLL resummation~\cite{hqt}. 
The two approaches agree very well in the kinematic range which is relevant 
for the envisaged experimental cuts.}
\label{fig:higgspt_nnlo}
\end{center}
\end{figure}
In~\cite{fehipWW} 
it was found that the average \pthiggs~of the Higgs boson, when discovery 
selection cuts are applied, can be as low as  $<\pthiggs> \simeq 15\; \GeV$. 
This leads to the question whether the NNLO result, unlike the NLO result, 
is a reliable prediction for such small values of the average \pthiggs. 
In Fig.~\ref{fig:higgspt_nnlo} we compare the integrated 
\pthiggs~distribution at NNLO against the resummed NNLL spectrum. 
We find a very good agreement between the two approaches for 
surprisingly low values of \pthmax. 
We conclude that higher than NNLO perturbative contributions, 
which  are accounted for with the resummation, remain small for the 
integrated \pthiggs~spectrum, when the maximum Higgs transverse momentum 
is restricted even down to $20 \,\GeV$.  

We have now validated the NNLO perturbative calculation~\cite{fehipWW} 
in a challenging case for fixed order perturbation theory which similarly 
emerges, due to the jet-veto and other cuts favoring 
small \pthiggs~values, in the search for 
a Higgs boson in the \WW~decay channel. 
However, the simulation of processes at NNLO 
is only performed at the parton level. We would like to investigate 
whether parton shower Monte-Carlo programs, which can also model 
non-perturbative effects and are computationally more flexible than 
NNLO Monte-Carlo's, 
provide realistic  estimates of the signal cross-section. 

We first discuss the problem of the normalization of the event generators.
Parton shower Monte-Carlo programs predict the same total cross-section 
as the cross-section for their encoded partonic hard scattering at fixed 
order in perturbation theory. Therefore, \HERWIG~predicts the 
Higgs boson total cross-section with LO accuracy (underestimating 
it by a factor of $\sim 2$) and \mcnlo~provides NLO precision (underestimating 
the total cross-section by a factor of $\sim 1.25$).  A matching of parton 
showers to NNLO fixed order calculations is not yet developed. 
Following a practical approach, we will validate whether 
the efficiency of experimental cuts and 
normalized differential distributions are in agreement with the NNLO 
calculations of Ref.~\cite{fehipWW}. 
We will then rescale the predictions of the 
\mcnlo~and \HERWIG~event generators with a global $K$-factor in 
order to  reproduce the fixed order result for the total 
cross-section. We will denote that the results of the 
Monte-Carlo ${\rm X}$  have been multiplied with a $K$-factor using the 
notation ${R(\rm X)}$.

\begin{figure}[h]
\begin{center}
\includegraphics[width=0.49\textwidth]{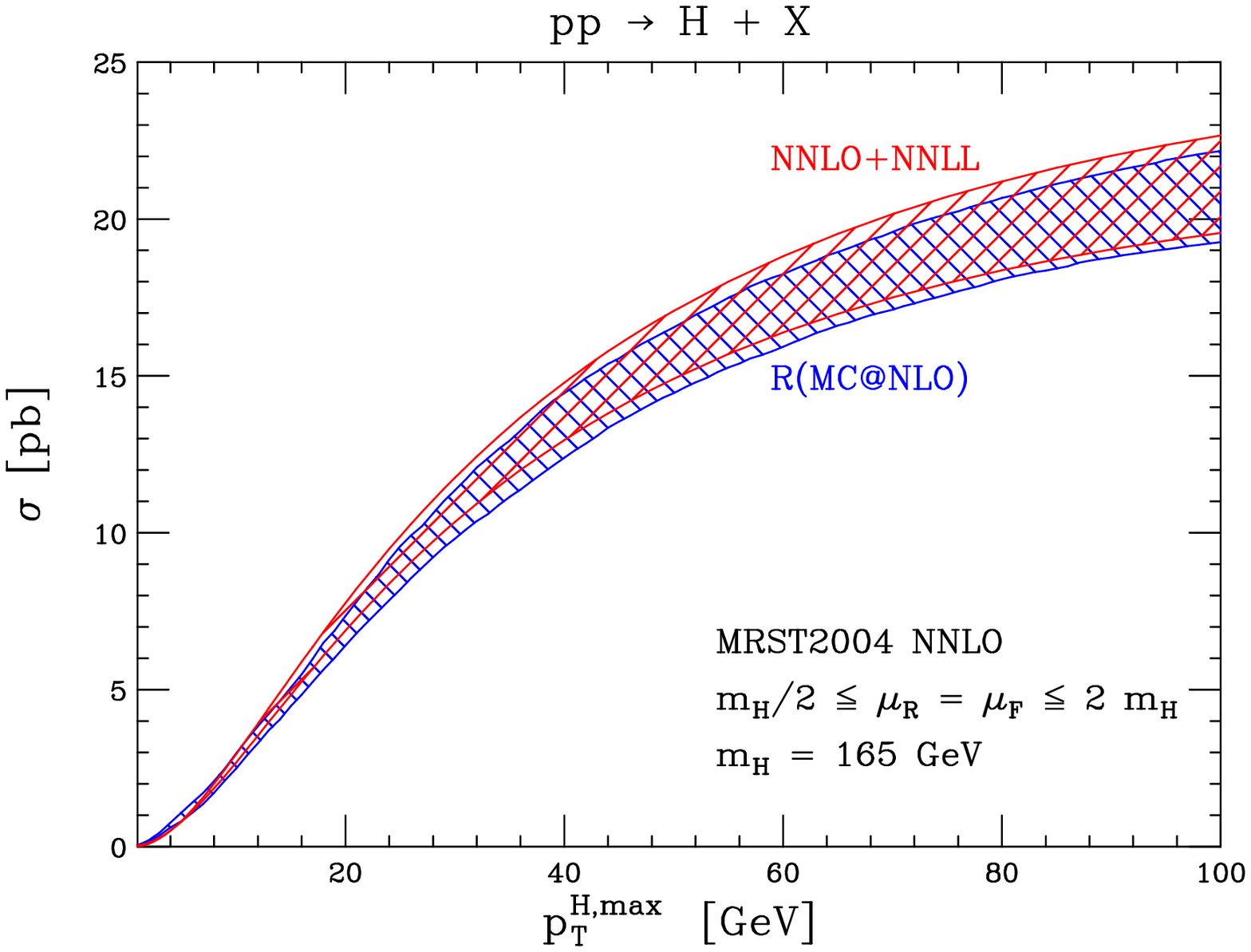}
\includegraphics[width=0.49\textwidth]{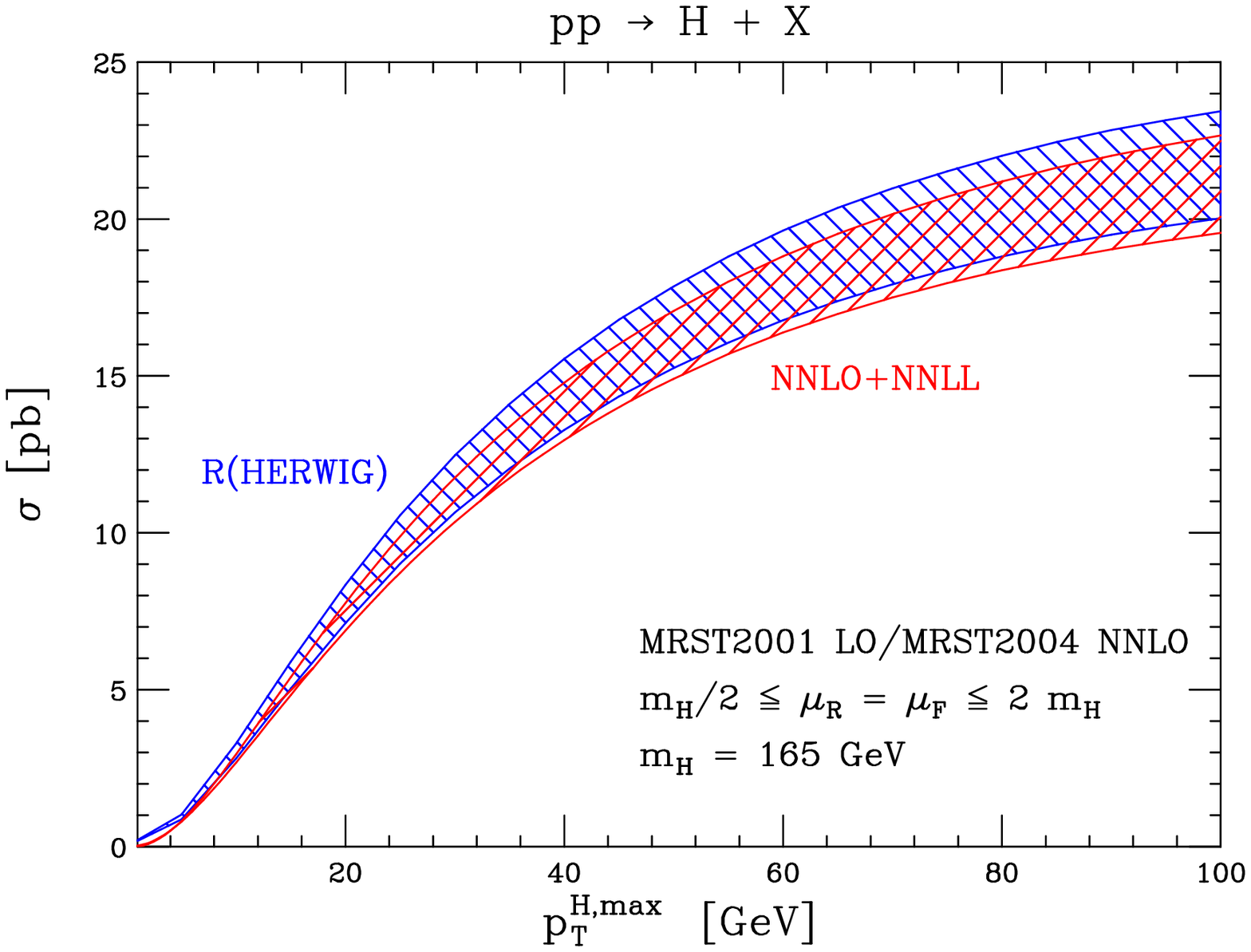}
\caption{Cumulative cross-section for the Higgs transverse momentum 
distribution. The scaled MC@NLO and \HERWIG~spectra agree very well 
with the resummed NNLL spectrum~\cite{hqt}.
}
\label{fig:higgspt_nnlo2}
\end{center}
\end{figure}
Now we will test how well event generators agree with resummation 
results for the \pthiggs~spectrum. 
In Fig.~\ref{fig:higgspt_nnlo2} we compare the integrated \pthiggs~
spectrum of \mcnlo~and \HERWIG~against the resummed NNLL prediction.    
We observe that both generators are in very 
good agreement with the NNLL spectrum. This is especially surprising for 
\HERWIG~which aims to describe the salient physics features of the 
process. 
Note, however, that \mcnlo~gives slightly larger and
\HERWIG~slightly smaller values than the NNLL resummation~\cite{hqt}. 

Before we conclude our analysis of the integrated \pthiggs~ distribution
we wish to comment further on the scale variation of the fixed order 
results. 
In Fig.~\ref{fig:higgspt_fixorder} we find a 
\pthmax~with no scale variation.  A similar behavior is 
also observed for 
the accepted cross-section with all experimental 
cuts~\cite{newcuts1,newcuts2} in Ref.~\cite{fehipWW}.  
It is therefore necessary to investigate in better detail the 
variation of the NNLO result. 
We  show the value of the cross-section at NNLO for $\pthmax \simeq 37 \,\GeV$,  varying independently the renormalization and factorization scales (the errors
correspond to the numerical integration):
\begin{center}
  \begin{tabular}{|l||cccc|} \hline
    $\sigma\;\;[pb]$   & $\mu_R=\frac{\MH}{4}$ & $\mu_R=\frac{\MH}{2}$ & $\mu_R=\MH$ & $\mu_R=2\MH$ \\\hline\hline
    $\mu_F=\frac{\MH}{4}$  &   $13.31\pm0.13$ & $13.76\pm0.08$ & $13.45\pm0.05$ & $12.82\pm0.04$ \\
    $\mu_F=\frac{\MH}{2}$  &   $13.15\pm0.13$ & $13.85\pm0.08$ & $13.69\pm0.06$ & $13.14\pm0.04$ \\
    $\mu_F=\MH$            &   $13.13\pm0.13$ & $14.00\pm0.08$ & $13.96\pm0.06$ & $13.47\pm0.04$ \\
    $\mu_F=2\MH$           &   $13.05\pm0.13$ & $14.15\pm0.08$ & $14.21\pm0.06$ & $13.76\pm0.04$ \\ \hline
  \end{tabular}
\end{center}
The cross-section is more sensitive to independent variations of  the 
renormalization and factorization scales; however, this variation 
is significantly smaller than in the total cross-section. A detailed study 
of the cross-section when all experimental discovery cuts are applied, which 
shows a similar scale-variation pattern,  
has been made in Ref.~\cite{fehipWW}.

Summarizing, in this Section we found that the integrated \pthiggs~spectrum is 
predicted reliably at NNLO for the kinematic range of \pthiggs~which is 
relevant in the search 
\ppHWWlept. On the contrary, the NLO fixed order 
calculation is unreliable.  We have also established that for the same 
observable, \mcnlo~and \HERWIG~are in a very good agreement with the 
resummed NNLL spectrum when they are normalized to a common NNLO total 
cross-section.

\section{Kinematic distributions and signal cross-section}
\label{sec:comp2}
The main backgrounds for the \ppHWWlept~
process are $\pptt$ and  $\ppWW$.  
These backgrounds are sufficiently suppressed to allow for the 
discovery of a Higgs boson  with  a combination of 
experimental cuts~\cite{newcuts1,newcuts2}, 
exploiting the spin-correlations in the decay of the 
Higgs boson and the high average transverse momentum of jets 
in top-pair events. In Ref.~\cite{fehipWW}, the signal cross-section with 
these cuts was computed at NNLO. 
In this section we will compare the NNLO results of Ref.~\cite{fehipWW} 
with \mcnlo. The public version of \mcnlo~includes only partial 
spin-correlations for the $\HWWlept$ decay. 
Here, the full spin-correlations for the decay of the Higgs boson have 
been implemented in \mcnlo.  All the results of this paper correspond 
to  a Standard Model Higgs boson mass of $\MH = 165 \, \GeV$.

\begin{figure}[h]
\begin{center}
\includegraphics[width=0.65\textwidth]{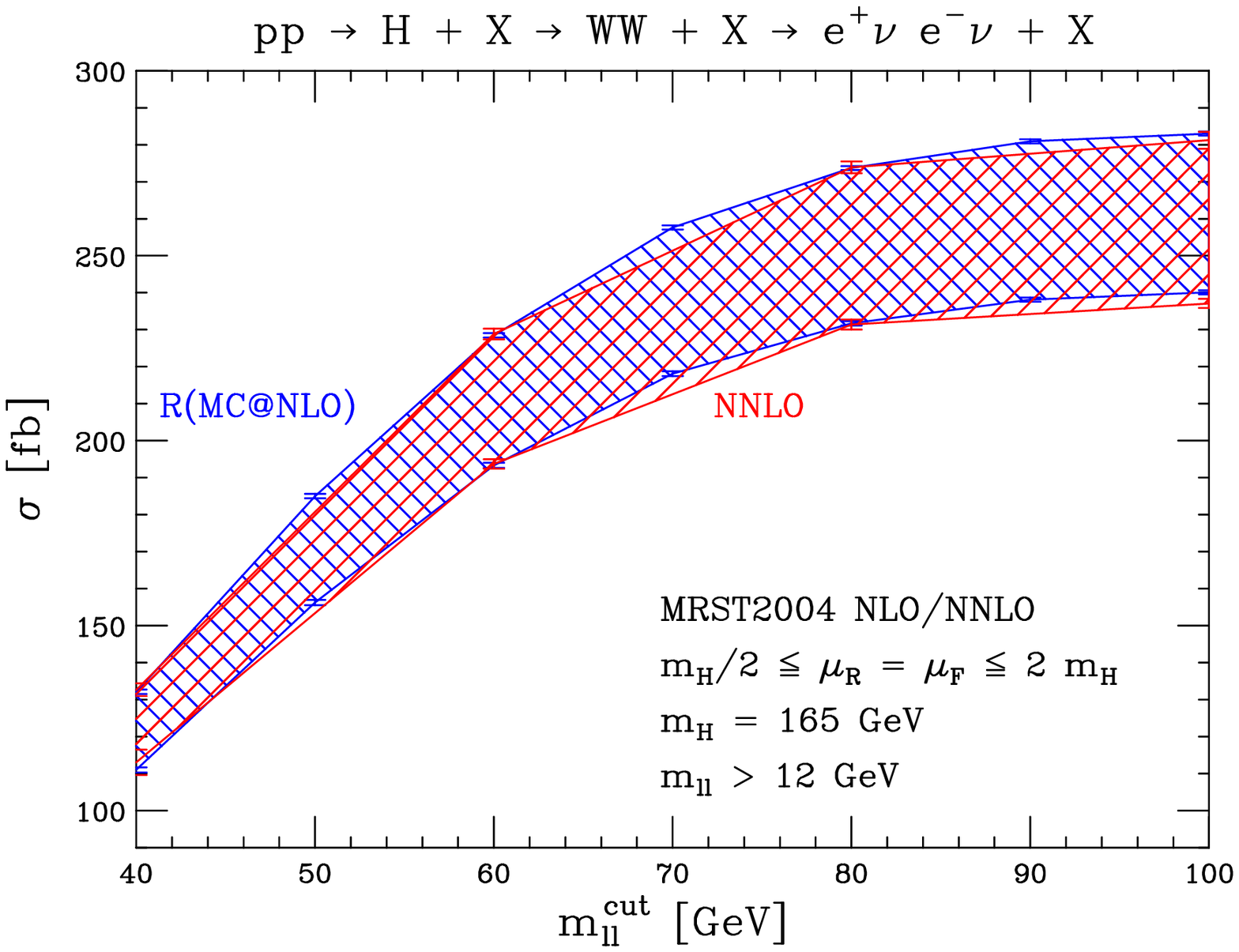}
\caption{Cross-section when the lepton invariant mass 
is constrained in the interval $\left[12\, \GeV,  \mll^{\mathrm{cut}}\right] $
at NNLO and with \mcnlo.
}
\label{fig:cut-mll}
\end{center}
\end{figure}

\begin{figure}[h]
\begin{center}
\includegraphics[width=0.65\textwidth]{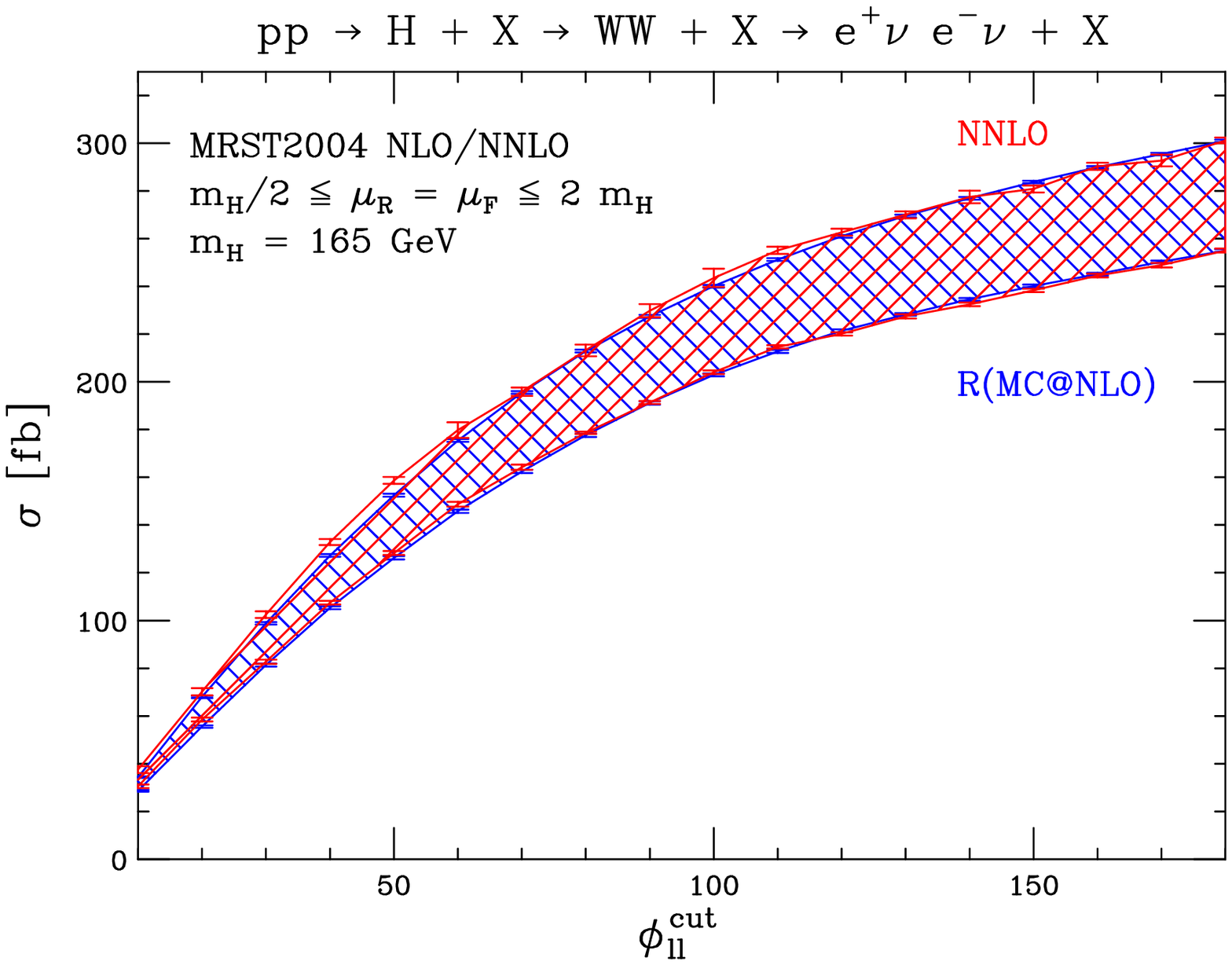}\\[0.3cm]
\caption{
Cross-section for the transverse opening angle of the two leptons in the 
interval  $\left[0,  \phill^{\mathrm{cut}}\right]$.
}
\label{fig:cut-phill}
\end{center}
\end{figure}

\begin{figure}[h]
\begin{center}
\includegraphics[width=0.65\textwidth]{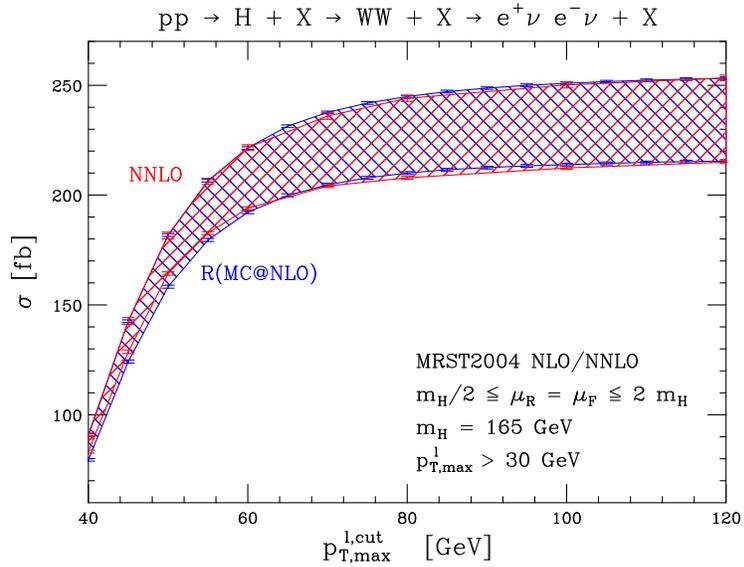}
\caption{ Cross-section when the maximum transverse momentum of the 
leptons is  in the range $\left[30\, \GeV,  \ptmaxcut \right] $. Each lepton should 
have a transverse momentum of at least $25\, \GeV$.}
\label{fig:cut-ptmax}
\end{center}
\end{figure}

\begin{figure}[h]
\begin{center}
\includegraphics[width=0.65\textwidth]{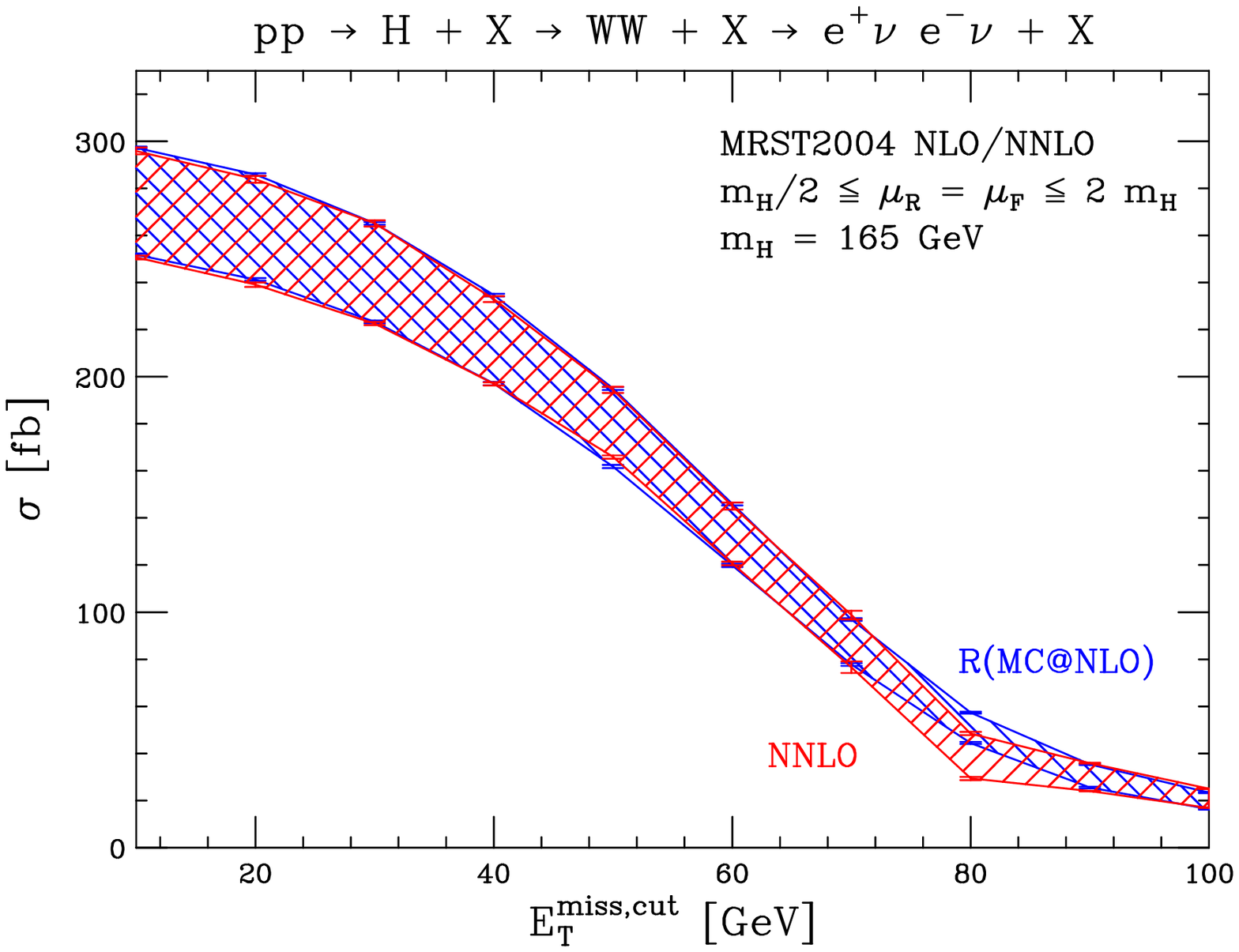}\\[0.3cm]
\caption{ Cross-section when the missing transverse energy is $\etmiss>\etmisscut$.}
\label{fig:cut-etmiss}
\end{center}
\end{figure}

In Figs.~\ref{fig:cut-mll}-\ref{fig:cut-etmiss} 
we present the cross-sections when a single cut is applied on
\begin{itemize}
\item $\mll$, the invariant mass of the charged lepton pair, 
\item $\phill$, the angle between the two charged leptons in the plane transverse to the beam axis, 
\item $\ptmax$, the transverse momentum of the harder lepton, and
\item $\etmiss$, the missing transverse energy.
%\item $\ptveto$, the maximum transverse momentum for jets with 
%$|\eta_\mathrm{jet}|<2.5$.
\end{itemize}
The four distributions  show an excellent agreement for the 
efficiencies among NNLO and \mcnlo. This is a remarkable result and 
could not have been easily foreseen; there is a 
significant change in the shape of the $\phill$, $\ptmax$, and $\etmiss$ 
distributions from NLO to NNLO, as seen in Ref.~\cite{fehipWW}.

A crucial experimental cut for suppressing the top-pair contribution to 
the background is a jet-veto. 
We veto events which have a transverse momentum of the leading jet in 
the central rapidity region ($|\eta_\mathrm{jet}|<2.5$) that is larger than \ptveto. 
For the jet definition we use here a \kt~algorithm 
with a jet-radius parameter $R =0.4$; 
later we will also use a cone algorithm (SISCone~\cite{siscone}). The  
two algorithms are identical for the LO and NLO calculation where 
only up to one parton can be present in the final state, if the 
same jet-radius $R$ is used. 
They differ, however, in the parton shower calculations (\mcnlo~and HERWIG) 
and at NNLO as more partons are generated. 
In the envisaged experimental analysis a jet-veto with 
a rather small value of $\ptveto \sim 25-40 \, \GeV$ is considered. We will 
investigate  whether  the NNLO and \mcnlo~predictions 
are consistent with each other for such small values of 
the jet-veto. 

The good agreement 
of the integrated \pthiggs~distribution between NNLO, \mcnlo~and NNLL 
resummation suggests that a good agreement between 
\mcnlo~and the NNLO cross-sections with a jet-veto may also hold. 
The jet-veto cross-section  should be qualitatively 
similar to  the cross-section with a cutoff on the \pthiggs~since 
at NLO the Higgs transverse momentum  corresponds exactly to 
the transverse momentum of the additional jet. However, 
the two cuts are not exactly 
the same and they compare only qualitatively. 
The jet-veto applies only at central rapidities; in addition, 
beyond NLO the \pthiggs~is not the same variable as the maximum 
transverse momentum of the jets. 
\begin{figure}[h]
\begin{center}
\includegraphics[width=0.65\textwidth]{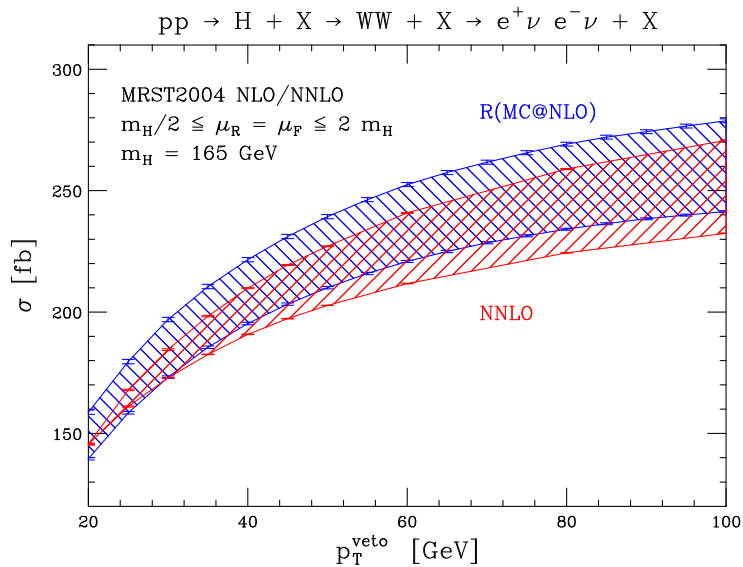}
\caption{The Higgs production cross-section with a 
fixed-order computation (NNLO) and  \mcnlo~rescaled with an 
inclusive $K$-factor (R(\mcnlo)) when a veto on jets with  $\pt > \ptveto$ 
at central rapidities $\left| \eta \right| < 2.5 $ is applied.
}
\label{fig:cut-jetveto}
\end{center}
\end{figure}
In Fig.~\ref{fig:cut-jetveto} 
we present the cross-section with a jet-veto applied.  
Indeed, we find a very good  agreement between the NNLO result 
and \mcnlo~(rescaled with the appropriate NNLO/NLO $K$-factor for 
the total cross-section). 

In Table~\ref{tab:acc-cross-sections} we list
the cross-section after all {\it signal cuts} as described in Ref.~\cite{fehipWW} are applied.
We have used both the $k_\mathrm{T}$ and SISCone algorithm of Ref.~\cite{siscone} and 
their implementation from Ref.~\cite{algos}. 
The jet radius in the azimuth-rapidity plane was set to $R=0.4$ and 
the merging parameter for the SISCone 
algorithm to $f=0.5$ \footnote{The merging parameter $f$ defines, how much two separate 
proto-jets need to overlap in order to be merged into one jet.}. The 
two algorithms  yield formally identical results for the fixed order 
calculation through NLO and indistinguishable results at NNLO 
within our Monte-Carlo integration precision~\footnote{We thank Gavin Salam
for pointing out to us that the SISCone and $\kt$ 
algorithms are formally different already at NNLO.}. 
In the first section of the Table we present the results obtained using a 
fixed-order LO computation and the LO+parton-shower 
event generator \HERWIG. We find a much larger fixed order result than 
when a parton shower is added. At fixed leading order all events have a Higgs 
boson with zero transverse momentum, and the jet-veto rejects none of the 
events. On the contrary, \HERWIG~generates a large fraction of 
events with $p_{\mathrm{T}}^{\mathrm{jet}} > \ptveto$.

\begin{table}[h]
\begin{center}
\begin{tabular}{|l|cc|cc|}
\hline
  $\accsigma$ [\fb]          & \multicolumn{2}{c|}{$\mu=\frac{\mh}{2}$} & \multicolumn{2}{c|}{$\mu=2\;\mh$}      \\
  jet algorithm              & SISCone & $\kt$ & SISCone & $\kt$ \\\hline\hline
  LO                         & \multicolumn{2}{c|}{$21.00\pm0.02$}      & \multicolumn{2}{c|}{$14.53\pm0.01$}    \\
  \HERWIG                      & $11.16\pm0.04$   & $11.59\pm0.04$  & $7.60\pm0.03$  & $7.89\pm0.03$     \\ \hline\hline
  NLO                        & \multicolumn{2}{c|}{$22.40\pm0.06$}      & \multicolumn{2}{c|}{$19.52\pm0.05$}    \\
  \mcnlo                     & $17.42\pm0.08$   & $18.42\pm0.08$  & $13.60\pm0.06$  & $14.39\pm0.06$  \\
  $R^{\mathrm{NLO}}$(\HERWIG)  & $19.79\pm0.07$   & $20.56\pm0.07$  & $14.61\pm0.05$ &$15.17\pm0.05$    \\\hline\hline
  NNLO                       
& $18.18 \pm 0.43$ & $18.45\pm0.54$      
& $18.76 \pm 0.31$ & $19.01\pm0.27$    
\\
  $R^{\mathrm{NNLO}}$(\mcnlo)& $19.33\pm0.09$   & $20.43\pm0.09$  & $17.24\pm0.07$  & $18.24\pm0.07$  \\
  $R^{\mathrm{NNLO}}$(\HERWIG) & $22.02\pm0.08$   & $22.88\pm0.08$  & $18.65\pm0.07$ & $19.38\pm0.07$   \\\hline
\end{tabular}
\caption{Cross-sections after the {\it signal cuts}  of Ref.~\cite{fehipWW} are 
applied for different calculation methods. The statistical 
integration errors are shown expicitly.  The \mcnlo~and \HERWIG~cross-sections 
are evaluated with 1,000,000 generated events. The fixed-order results 
were computed in Ref.~\cite{fehipWW} and require the Monte-Carlo integration 
of multiple sectors~\cite{fehip}.}
\label{tab:acc-cross-sections}
\end{center}
\end{table} 

In the second section of Table~\ref{tab:acc-cross-sections} we present the 
results obtained using a fixed-order NLO computation, the event generator 
\mcnlo~and \HERWIG~after we have rescaled it with an inclusive NLO/LO factor. 
While in \mcnlo~and fixed order (LO, NLO and NNLO)
we can set the renormalization and factorization scales for the 
hard scattering at will, we use \HERWIG~at the default scale since this 
affects the triggering of the hadronization procedure. We then rescale 
the \HERWIG~result using a $K$-factor, 
taking the NLO fixed order result to be at
the scale $\mu = \MH/2$ or $\mu = 2 \MH$.
The NLO result is quite different from the one obtained with \mcnlo. 
We can attribute this failure of the NLO computation to the 
poor modeling of the low $\pt$ region, as is shown in 
Fig~\ref{fig:higgspt_nlo}.
 
In the last part of Table~\ref{tab:acc-cross-sections} we present the 
results obtained using a fixed-order NNLO computation and the results
from  \mcnlo~and \HERWIG, rescaled to the NNLO total 
cross-section.
The NNLO result and the rescaled \mcnlo~give consistent results, 
albeit with different behavior when varying the renormalization and
factorization scales. A detailed analysis of the NNLO scale dependence when 
all cuts are applied can be found in Ref.~\cite{fehipWW}.

We note that for all the results  of this section  we use 
\mcnlo~and \HERWIG~at the parton level, switching off the hadronization 
and without using  a model for the underlying event.
We observe that the \kt-algorithm gives larger cross-sections 
than the SISCone algorithm for \mcnlo~and \HERWIG; 
as mentioned before, the results for the two algorithms 
are very similar at fixed order through NNLO 
(within our integration precision). 
Additionally \mcnlo~gives slightly 
smaller values for the cross-sections than \HERWIG.  

We have established in this section that the efficiency of experimental cuts 
computed with \mcnlo~and \HERWIG~is similar to the efficiency obtained 
at NNLO.  There are rather dramatic changes in differential distributions 
when going from NLO to NNLO~\cite{fehipWW}.  It is only at NNLO that the 
fixed order calculation is consistent with the parton shower 
efficiency of the experimental cuts. 
In the following Section we will study the dependence of the 
cross-section on effects that are not captured by the fixed order 
NNLO calculation. We will study the effect of hadronization and of the 
underlying event.  We will also investigate further the differences in the 
cross-section due to the two different jet algorithms.

\section{Jet algorithms, hadronization and the underlying event}
\label{sec:jetalgo}
In this Section we perform a study of the signal cross-section 
with all cuts applied using \mcnlo.  
We will analyze the impact of different 
jet clustering methods, hadronization 
and the underlying event. 

\begin{figure}[h!]
\begin{center}
\includegraphics[width=0.65\textwidth]{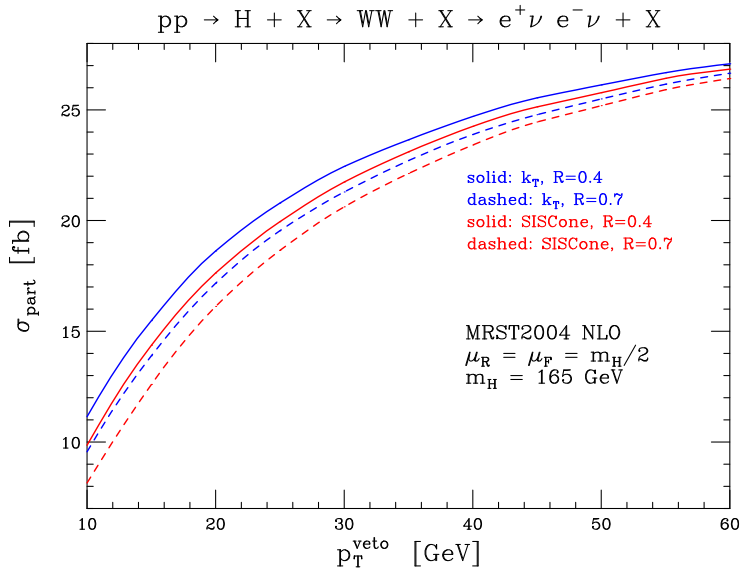}
\caption{Comparison between the cone and \kt~algorithm for different 
values of the allowed maximum jet transverse energy (jet-veto). All other 
cuts are set to the values chosen in Ref.~\cite{fehipWW} for the 
{\it signal cuts}.}
\label{fig:jetalgo-ptveto}
\end{center}
\end{figure}

In Fig.~\ref{fig:jetalgo-ptveto} we plot the cross-section using 
\mcnlo~as a function of the  \ptveto~value for the 
\kt~and the SISCone algorithm with a jet-radius $R=0.4$ and $R=0.7$.  
The clustering  is applied to all final state particles before 
hadronization.
We find that for small values of the jet-veto parameter the choice of 
the jet clustering method is more significant. 
For a jet-veto at $\ptveto = 25 \, \GeV$ the choice of  jet-algorithm
changes the cross-section by $\sim 6\%$ with \mcnlo. 
A similarly large variation of $\sim 7\%$ is observed 
when we vary the jet-radius  from $R = 0.4$ to $R=0.7$.
For a jet veto value larger than 
about $\ptveto \simeq 40 \, \GeV$ the sensitivity 
of the cross-section to the choice of the jet-algorithm 
or the jet-radius falls below $\sim 2-3\%$.

\begin{figure}[h!]
\begin{center}
\includegraphics[width=0.49\textwidth]{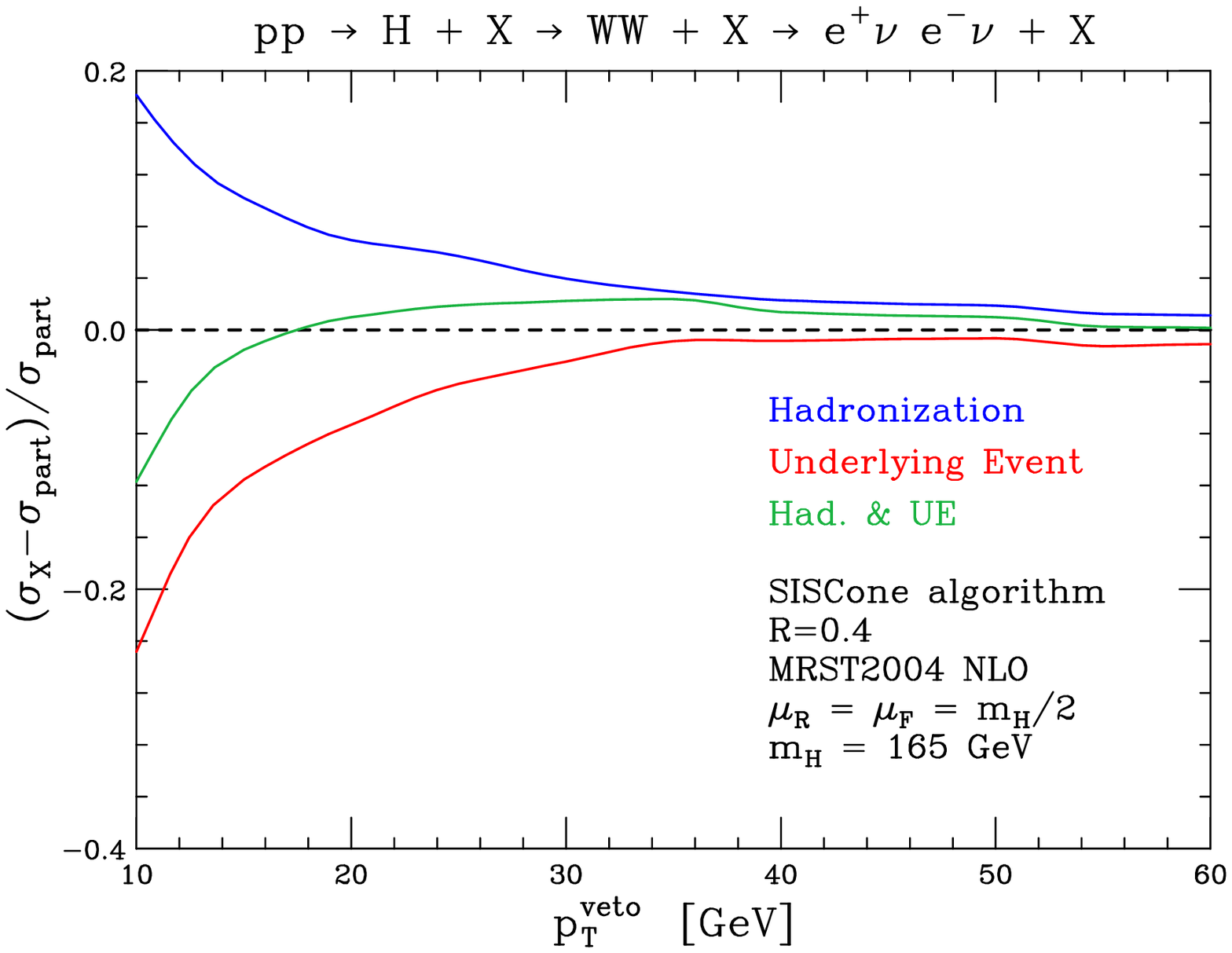}
\includegraphics[width=0.49\textwidth]{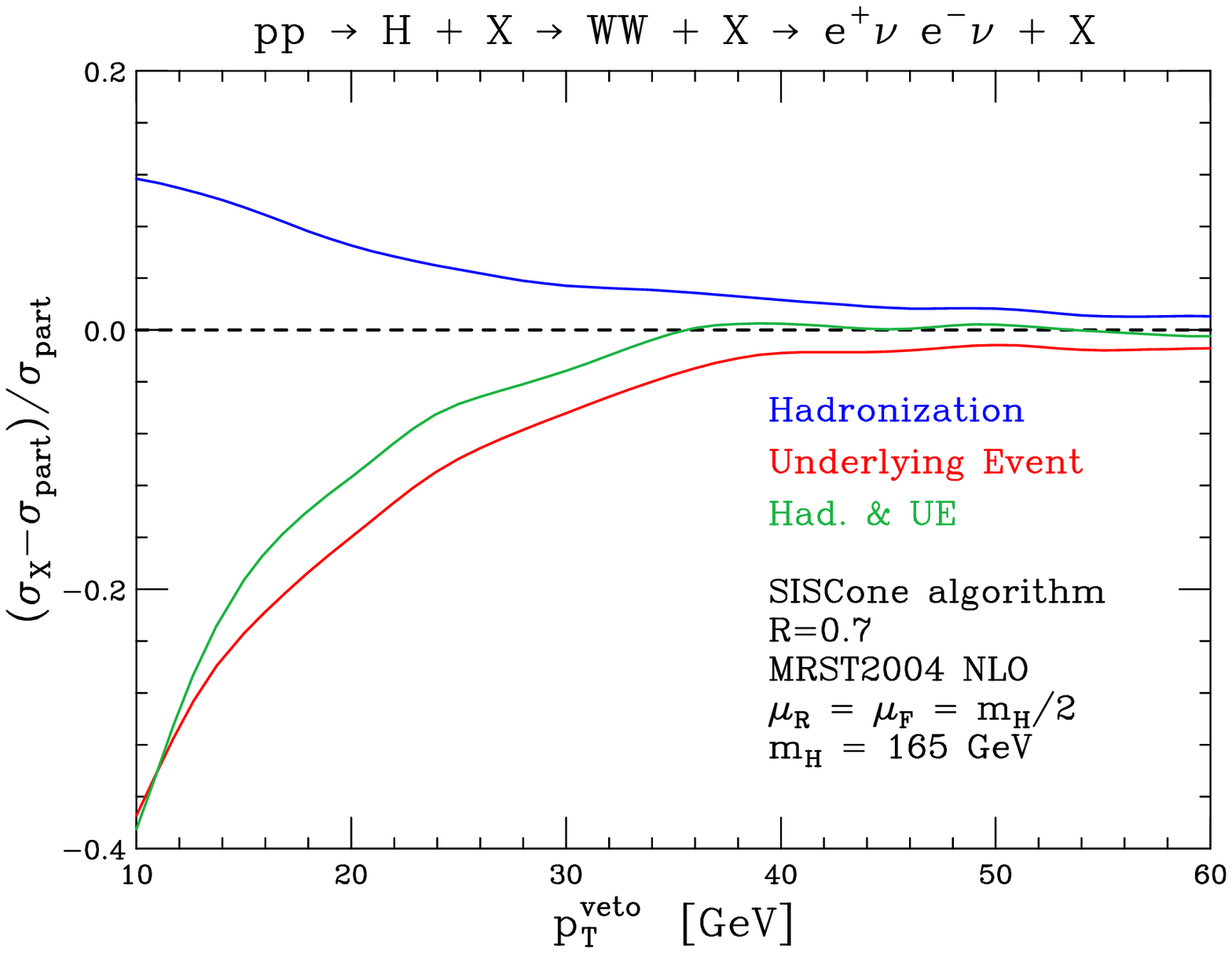}\\
\caption{Difference of the cross-section after {\it signal cuts} including the underlying event and 
hadronization models, with respect to the partonic cross-section. 
The cross-section is shown as a function 
of the jet-veto value for the SISCone clustering algorithm.}
\label{fig:jetalgo-ptveto-SISCone}
\end{center}
\end{figure}

We now study the effect of hadronization as it is modeled in 
\HERWIG~and of the underlying event as implemented  in  
JIMMY~\cite{jimmy}. In Fig.~\ref{fig:jetalgo-ptveto-SISCone} 
we present the relative difference of the 
cross-section with respect to 
the partonic cross-section 
when the hadronization or/and the 
underlying event models are switched on. 
We have used here the SISCone algorithm with a merging 
parameter $f = 0.5$ and two values for the jet-radius 
R=0.4 (left) and R=0.7 (right). 
We apply the {\it signal cuts} set to the values 
which are used in Ref.~\cite{fehipWW}. We vary, however,
the allowed maximum value of \ptjet.  
Of interest are values of the jet-veto between 25 and 
40 \GeV, which are envisaged in the Higgs boson search.

Qualitatively, we anticipate that the 
hadronization and the underlying event change 
the partonic cross-section with opposite signs 
(we refer the reader to the recent analysis in 
Ref.~\cite{underevent} for a detailed study).  
Hadronization reduces the average $\pt$  of (gluonic) jets by 
roughly $\delta \pt \sim (1 \, \GeV)/R$. 
The underlying event increases
the jet \pt~by  roughly $\delta \pt \sim R^2\times (5 \, \GeV)$ 
at the LHC. The slope of the partonic 
cross-section with the jet-veto cutoffs 
(Fig.~\ref{fig:cut-jetveto}) is large for small values of 
the jet-veto. The shifts $\delta \pt$ from hadronization and 
the underlying event can therefore induce significant changes 
to the cross-section. A jet-veto 
after hadronization corresponds to a looser effective 
jet-veto at the parton level. We therefore anticipate 
the cross-section to increase by switching on the 
hadronization model. Similarly, we anticipate a decrease of 
the cross-section due to the underlying event. 

The trends can be verified in 
Fig.~\ref{fig:jetalgo-ptveto-SISCone}. A smaller jet-radius 
increases the impact of hadronization and decreases the 
impact of the underlying event. The two effects are not 
linearly additive.  However, we find that a cancelation between
the two effects, which varies according to the jet-radius,
takes place.  For a jet veto $\ptveto = 25 \, \GeV$ 
and a radius $R = 0.4$, the hadronization shift is 
about  $\sim 7\%$ and the underlying event shift is 
$\sim 4\%$. For a larger radius $R = 0.7$, the two 
shifts are $5\%$ and $10\%$ correspondingly.

\begin{figure}[h!]
\begin{center}
\includegraphics[width=0.49\textwidth]{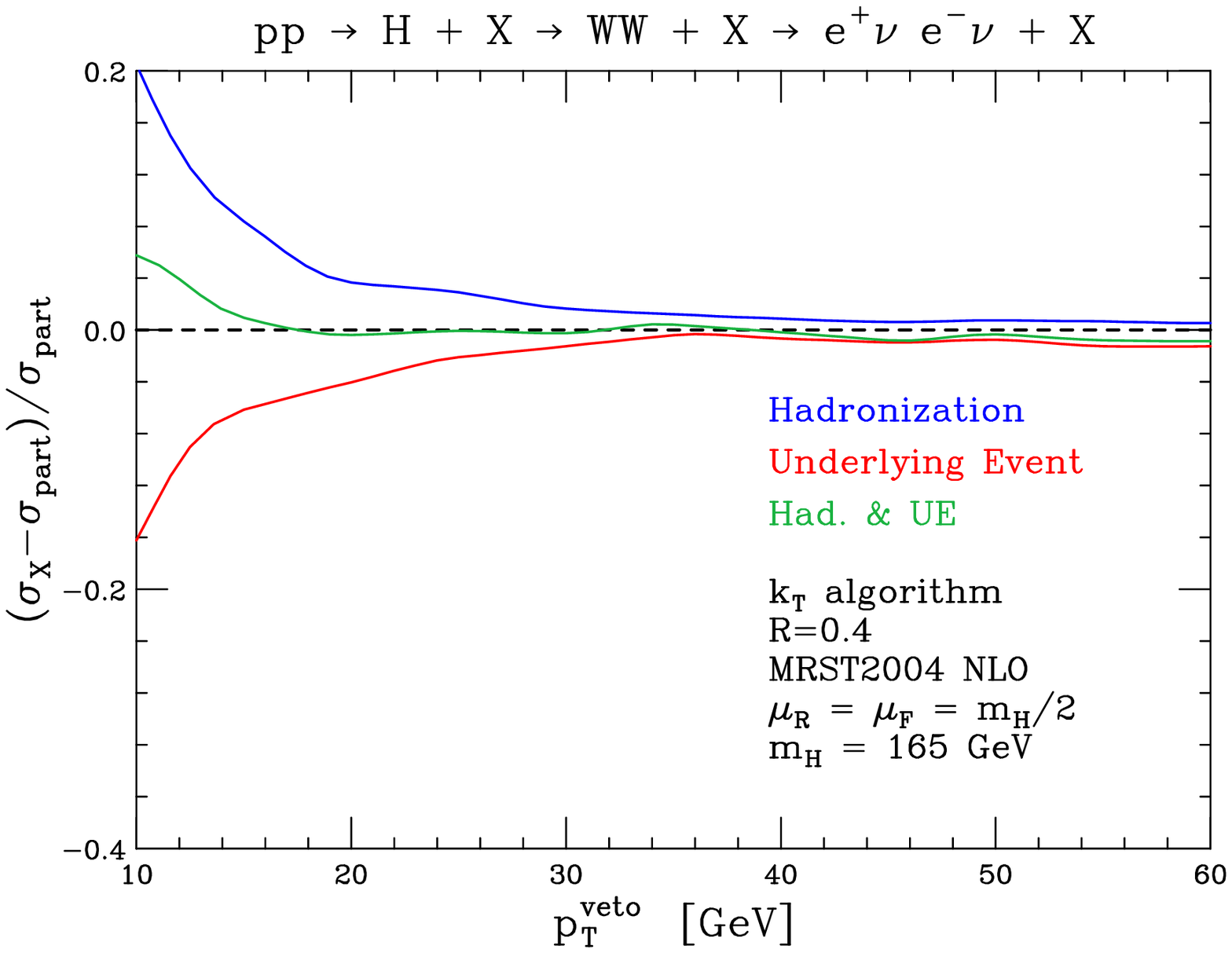}
\includegraphics[width=0.49\textwidth]{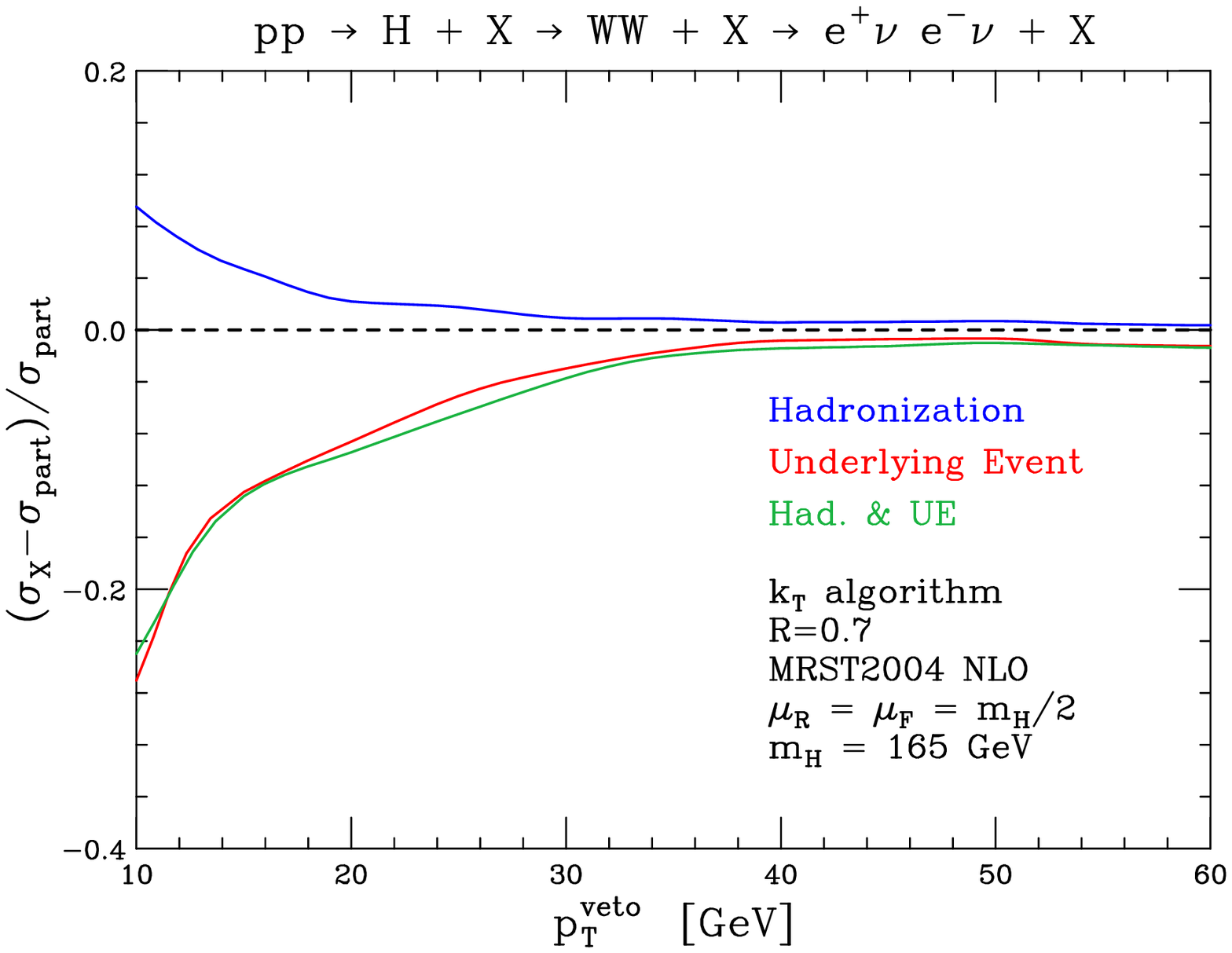}\\
\caption{Difference of the cross-section after {\it signal cuts} 
including the underlying event and 
hadronization models, with respect to the partonic cross-section. 
The cross-section is shown as a function 
of the jet-veto value for the \kt~clustering algorithm.}
\label{fig:jetalgo-ptveto-Kt}
\end{center}
\end{figure}

In Fig.~\ref{fig:jetalgo-ptveto-Kt} we show 
the effects of hadronization and the underlying event 
for the \kt~algorithm. We find the same features qualitatively 
for the two effects as in the SISCone algorithm. We note however, 
that the \kt-algorithm shows an overall reduced sensitivity.

\section{Conclusions}
\label{sec:TheEnd}

In this paper we have studied QCD effects for the process 
$\ppHWWlept$. This process is of particular 
interest at the LHC since it is possible that for a range of 
mass values of the Higgs boson, this channel is the only viable 
one for a discovery.

The cross-section with all envisaged experimental cuts applied  
was  computed in an earlier publication~\cite{fehipWW} at NNLO in 
QCD. In this paper we compared these NNLO results with the 
leading order event generator \HERWIG~\cite{herwig} and the  
event generator \mcnlo~\cite{mcnlo} which performs a matching of 
\HERWIG\ with NLO fixed order perturbation theory. We found very good 
agreement in efficiencies  of all experimental cuts that are relevant 
in the search for the Higgs boson.  
This is rather spectacular given that there  are 
significant corrections in the total cross-section and the shape of 
kinematic distributions from NLO to NNLO.  

The experimental cuts select events with small transverse momentum 
of the Higgs boson. This is important in order to reduce the selection 
of events from top-quark production, which is a major background. 
We have compared 
a NNLO computation and the result of NNLL resummation~\cite{hqt} for the 
cumulative $\pthiggs$ distribution.  We found that NNLL resummation  does 
not induce significant corrections with respect to the NNLO calculation 
for the kinematic range which is favoured  by the selection cuts. 
We have also found that, within the uncertainty from scale variations, 
\mcnlo~and \HERWIG~  are in very good agreement with the NNLL result.
On the contrary, fixed order NLO perturbation theory provides a rather 
poor approximation for the required distributions and efficiencies.

Finally, we investigated the magnitude of effects due to
hadronization, the underlying event, and the choice of 
jet algorithms. For typical choices of parameters and cutoffs 
in the  experimental cuts  we find a mild dependence of the 
cross-section  on these effects.  
In this paper we did not examine a variety of 
remaining uncertainties, such as  uncertainties in 
the parton densities,  which may  be relevant at a precision 
level of 10\%.  Studies  of electroweak corrections  can be found 
in Refs.~\cite{ewk1,ewk2,ewk3} and of  
the background-signal interference in Ref.~\cite{BS}.

The results and comparisons made in this paper provide 
a firm validation of the fixed order NNLO results 
and the event generator tools which are available for 
simulating the $\ppHWWlept$ 
process at the LHC.

\section*{Acknowledgements}
We thank Michael Dittmar, Stefano Frixione, Massimiliano Grazzini and 
Giulia Zanderighi for useful discussions and suggestions. 
This research was supported by the Swiss 
National Science Foundation  under contracts 200021-117873 
and 200020-113378/1.\\ 
We thank Gavin Salam
for pointing out to us that the SISCone and $\kt$ 
algorithms are formally different at NNLO.

%%%%%%%%%%%%%%%%%%%%%%%%%%%%%%%%%%%%%%%%%%%%%%%%%
%%%%% bibliography
%%%%%%%%%%%%%%%%%%%%%%%%%%%%%%%%%%%%%%%%%%%%%%%%%%

%\newpage

\bibliographystyle{JHEP}

\end{document}